\documentclass[12pt,preprint]{aastex}
\usepackage{epsf}
\usepackage{graphicx}

\newcommand{\oiii}{[O\,{\sc iii}]}

\newcommand{\neiii}{[Ne\,{\sc iii}]}
\newcommand{\neiv}{[Ne\,{\sc iv}]}
\newcommand{\nev}{[Ne\,{\sc v}]}
\newcommand{\ariv}{[Ar\,{\sc iv}]}
\newcommand{\hb}{H$\beta$}
\newcommand{\fiv}{[F\,{\sc iv}]}

\newcommand{\kms}{km s$^{-1}$}

\received{}
\revised{}
\accepted{}

\slugcomment{}
\shorttitle{Detection of Fluorine in the Halo PN BoBn 1}
\shortauthors{Otsuka et al.}

\begin{document}

\title{Detection of Fluorine in the Halo Planetary Nebula
BoBn 1: \\Evidence For a Binary Progenitor Star}
\author{Masaaki {\sc Otsuka}\altaffilmark{1,2}, 
Hideyuki {\sc Izumiura}\altaffilmark{1}, Akito {\sc Tajitsu}\altaffilmark{3}, and 
Siek {\sc Hyung}\altaffilmark{4}}

\altaffiltext{1}{Okayama Astrophysical Observatory (OAO), NAOJ, 
Kamogata, Okayama 719-0232, Japan; 
masaaki@oao.nao.ac.jp, izumiura@oao.nao.ac.jp.}
\altaffiltext{2}{Current address: Space Telescope Science
Institute, 3700 San Martin Drive, Baltimore, MD 21218, U.S.A.; otsuka@stsci.edu.}
\altaffiltext{3}{Subaru Telescope, NAOJ, 650 North A'ohoku Place, Hilo,
Hawaii 96720, U.S.A.; tajitsu@subaru.naoj.org.}
\altaffiltext{4}{School of Science Education (Astronomy), Chungbuk National
University, 12 Gaeshin-dong Heungduk-gu, CheongJu, Chungbuk 361-763,
Korea; hyung@chungbuk.ac.kr.}
\begin{abstract}
We have found the fluorine lines {\fiv} $\lambda\lambda$3996.92,4059.90 
in the extremely metal-poor ([Ar/H] = $-$2.10$\pm$0.21) halo planetary nebula 
(PN) BoBn 1 in high-dispersion spectra from the 8.2-m VLT/UVES archive. 
Chemical abundance analysis 
shows that the fluorine abundance is 
[F/H] = +1.06$\pm$0.08, making BoBn 1 the most fluorine-enhanced and
 metal-poor PN among fluorine-detected PNe and providing new evidence 
that fluorine is enhanced by nucleosynthesis in low mass metal-poor stars. 
A comparison with the abundances of carbon-enhanced metal-poor (CEMP) 
stars suggests that BoBn 1 shares their origin and evolution with 
CEMP-{\it s} stars such as HE1305+0132. BoBn 1 might have evolved from 
a binary consisting of $\sim$2 M$_{\odot}$ primary and 
$\sim$0.8 M$_{\odot}$ secondary stars. 
\end{abstract}
\keywords{ISM: abundances -- planetary nebulae: individual (BoBn 1)}

\section{Introduction}
Currently, over 1,000 objects are regarded as planetary nebulae (PNe) 
in the Galaxy (Acker et al. 1992), 
while 14 of them have been identified as halo members from their
location and kinematics since the discovery of K 648 (Ps 1) in M 15. 
All halo PNe are metal-poor 
objects ($\langle$[S/H]$\rangle$=$-$1.3; Pereira \& Miranda 2007) 
and are suspected to be descendants of stars formed in the early history 
of the Galaxy. Halo PNe should convey important information for 
the study of the evolution of metal-poor stars as well as the early
physical and chemical conditions of the Galaxy. Their origins are,
however, not well understood yet.

BoBn 1 (PN G108.4--76.1) is a halo PN that is located in 
the direction of the south Galactic pole. The heliocentric distance 
to this object is estimated to be between 16.5 kpc (Henry et al. 2004) 
and 29 kpc (Kingsburgh \& Barlow 1992) and the heliocentric radial 
velocity is +191.9$\pm$0.8 {\kms} (Otsuka 2007). Zijlstra et al. (2006) 
have associated BoBn 1 with the leading tail of the Sagittarius Dwarf 
Spheroidal Galaxy, which traces several halo globular clusters. BoBn 1 
is an extremely metal-poor ([Ar/H] = $-$2.10$\pm$0.21; Otsuka 2007) and C- and 
N-rich halo PN ([C/O] = +1.50$\pm$0.27, [N/O] = +1.02$\pm$0.13, 
[O/H] = $-$0.96$\pm$0.05, according to Otsuka 2007; 
[C/O] = +1.32, [N/O] = +1.07, [O/H] = $-$1.07, according to 
Howard et al. 1997), which composes a class of PN together 
with K 648 and H 4-1. 

These C- and N-rich halo
PNe have some unresolved issues on their origin. For example, halo PNe
can become N-rich, 
 but not C-rich, if the initial mass of the progenitor is $\sim$0.8
M$_{\odot}$, which is the turn-off star mass of M 15, 
according to the theoretical models of single metal-poor stars 
(e.g., Fujimoto et al. 2000). To become C-rich PNe, the third dredge-up 
(TDU) must take place in the late asymptotic giant branch (AGB) phase. 
But, the TDU takes place in stars with initial masses $\gtrsim$ 1.2--1.5 
M$_{\odot}$
(e.g., Straniero et al. 2006). Also, current stellar evolutionary models 
predict that the post-AGB evolution of stars with initial masses 
$\sim$0.8 M$_{\odot}$ proceeds too slowly for a visible PN to be formed.
These issues must be resolved before one proceeds to discuss the 
evolution of metal-poor stars and the early chemical evolution of 
the Galaxy through the observation of halo PNe. 

Chemical abundance analysis of PNe is a key tool to investigate these issues. 
Comparing the observed abundances with those from theoretical evolution 
and nucleosynthesis models for metal-poor stars will shed light
on the origin and evolution of C- and N-rich halo PNe.
In this paper, we present the first estimation of the fluorine (F)
abundance of BoBn 1. The abundances of F and heavy elements via the slow 
neutron capture process (the {\it s}-process), which are synthesized 
in the He-rich intershell of AGB stars and are brought to the surface by 
the TDU, would be an effective tool to examine how 
BoBn 1 became C- and N-rich, when combined with the C abundance. 
We discuss the origin and evolution of BoBn 
1 by comparing the F and other elemental abundances with the abundances 
of carbon-enhanced metal-poor (CEMP) stars.

\section{Data \& Reduction} 
High-dispersion spectra of BoBn 1 in the range 3260 {\AA} to 6680
{\AA} are available from the European 
Southern Observatory (ESO) archive.
The observations were performed by Perinotto \& Gilmozzi (Prop. I.D.:
69.D-0413A) on August 4, 2002, using the
Ultraviolet Visual Echelle Spectrograph (UVES; Dekker et al. 2000)
at the Nasmyth B focus of KUEYEN, the second of four 8.2-m telescopes
of the ESO Very Large Telescope (VLT) at Paranal, Chile. The entrance slit 
size was 11$\farcs0$ in length and 1$\farcs5$
in width, giving a spectral resolution of $\sim$36,000. 
The sampling pitch along the wavelength 
dispersion was $\sim$0.02 {\AA} per pixel. 
The data were taken as a series of 2700-sec exposures and the total
exposure time was 10,800 sec. The seeing was between 
1$\arcsec$ and 1$\farcs5$ during the exposure. 
The standard star HR 9087 was observed for flux
calibration. Data reduction was performed mostly by the Image
Reduction and Analysis Facility (IRAF) software package distributed by
the National Optical Astronomy Observatory (NOAO).

\section{Results} 

\begin{figure}
\epsscale{0.9}
\plotone{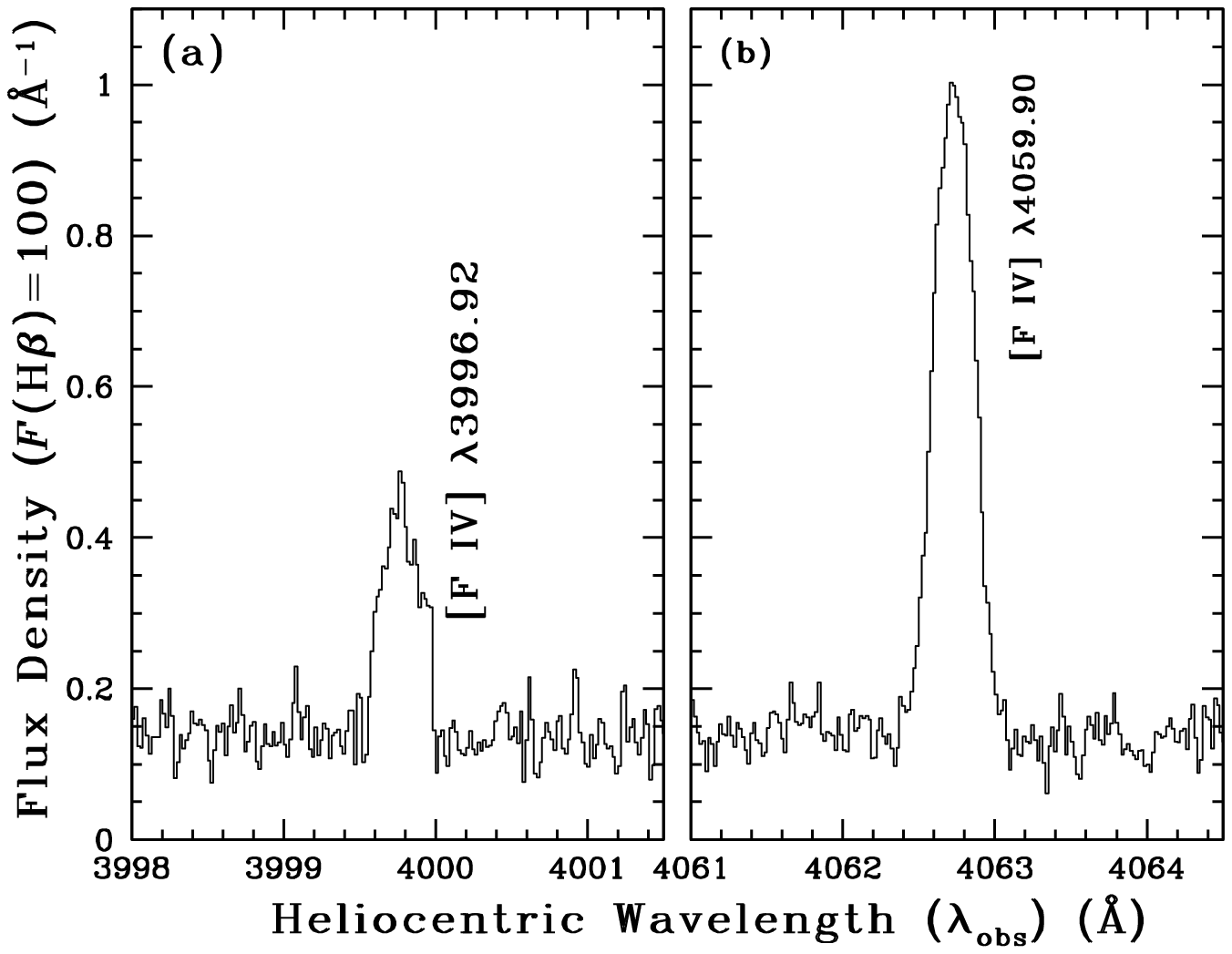}
\figcaption{Emission line-profiles of 
{\fiv} $\lambda$3996.92 ({\it left}) and {\fiv} 
$\lambda$4059.90 ({\it right}) of BoBn 1. \label{spec}}
\end{figure}

We have found candidates of the fluorine forbidden lines {\fiv} 
$\lambda$3996.92 (transition: $^{1}\rm D_{2}$--$^{3}\rm P_{1}$) and 
{\fiv} $\lambda$4059.90 ($^{1}\rm D_{2}$--$^{3}\rm P_{2}$). 
We show the line profiles of {\fiv} in 
Figure \ref{spec}.
The abscissa and ordinate axes indicate the heliocentric wavelength and 
the flux density  normalized to the total {\hb} flux $F$({\hb}). 
We have fitted the line-profiles of these {\fiv} with a single-Gaussian. 
The resultant fittings are summarized in Table
\ref{fl}. The second and third columns are the position of the center of the 
Gaussian and the normalized line intensity (interstellar reddening 
corrected using $c$({\hb}) = 0.21$\pm$0.02; Otsuka 2007), 
respectively. We confirmed that BoBn 1 has no high-density components
exceeding a critical density of 3$\times$10$^{6}$ 
cm$^{-3}$ at the level of $^{1}\rm D_{2}$, using the 
intensity ratios of Balmer lines (H{\it n}; {\it n} (principal quantum
number) = 3 -- 25) to H$\beta$ (H4), which are sensitive to high electron 
density ($>$ 10$^{4}$ cm$^{-3}$). Therefore, the effect of 
collisional de-excitation is negligibly small, so that the intensity
ratio of {\fiv} $I$($\lambda$4059.90)/$I$($\lambda$3996.92) is 
proportional to the ratio of their transition probabilities. When 
the transition probabilities $A$($^{1}\rm D_{2}$--$^{3}\rm
P_{1}$) = 3.42$\times$10
$^{-2}$ s$^{-1}$ and $A$($^{1}\rm D_{2}$--$^{3}\rm P_{2}$) = 
9.81$\times$10$^{-2}$ s$^{-1}$ given by Garstang (1951) -- they are in good
agreement with Storey \& Zeippen (2000) -- are adopted, 
$I$($\lambda$4059.90)/$I$($\lambda$3996.92) 
= 2.82, which is in good agreement with our measurement (3.01$\pm$0.40). 
Hence, these two emission
lines can be identified as the fluorine forbidden lines {\fiv} 
$\lambda$3996.92 and {\fiv} $\lambda$4059.90.

The triply ionized fluorine abundance F$^{3+}$/H$^{+}$ 
has been estimated from the each detected {\fiv}
line by solving the statistical
equilibrium equations for the lowest five energy levels ($^{3}\rm
P_{0,1,2}$, $^{1}\rm D_{2}$, $^{1}\rm S_{0}$). 
The collision strength and transition
probabilities of {\fiv} lines are given by Lennon 
\& Burke (1994) and Garstang
(1951), respectively.
To estimate F$^{3+}$/H$^{+}$, 
we used an electron temperature of $T_{\rm e}$ = 13,430$\pm$170 K,
derived from the ratio of {\oiii} 
$I$($\lambda$4959+$\lambda$5007)/$I$($\lambda$4363) 
= 80.05$\pm$2.67, and a density of $n_{\rm e}$ = 
4550$\pm$1270 cm$^{-3}$ from the
ratio of {\ariv} $I$($\lambda$4711)/$I$($\lambda$4740) = 1.01$\pm$0.08 
(Otsuka 2007). The resultant F$^{3+}$/H$^{+}$ 
are presented in the fourth column of Table \ref{fl}, which 
include the errors of the line
intensity, $T_{\rm e}$, $n_{\rm e}$, and $c$({\hb}). Finally, we adopted
a value of F$^{3+}$/H$^{+}$ from the intensity of the weighted mean, presented 
in the fourth column, last line of Table \ref{fl}. 

To estimate the elemental fluorine abundance F/H using only F$^{3+}$/H$^{+}$, 
we must correct for unobserved fluorine ionic abundances using an 
ionization correction factor icf(F). In BoBn 1, 
{\neiii}, {\neiv}, and {\nev} lines have been detected (Otsuka 2007). 
Using the similarity of the
ionization potential ranges for F$^{3+}$ (35--62.7 eV) and Ne$^{3+}$ (41--63.5
eV), F/H of BoBn 1 may be estimated by the following equations: 

\begin{eqnarray}
\rm F/H~~ &=&~~ \rm icf(F)~\times~\left(F^{3+}/H^{+}\right),\\
\rm icf(F)~~ &=&~~ \rm \left(Ne/H\right)~\times~\left(Ne^{3+}/H^{+}\right)^{-1},
\end{eqnarray}

\noindent where Ne/H and Ne$^{3+}$/H$^{+}$ are the elemental and triply 
ionized neon abundances, respectively. We assume that Ne/H of BoBn 1 
is the sum of Ne$^{2+}$/H$^{+}$, Ne$^{3+}$/H$^{+}$, and
Ne$^{4+}$/H$^{+}$. Finally, we estimated F/H as presented in the 
sixth column, last line of Table \ref{fl}. 

In Table \ref{fl2}, 
we present the [F/H] abundance of 11 disk PNe and a halo PN NGC 4361 from 
Zhang \& Liu (2005) and of NGC 7662 from Hyung \& Aller (1997), 
together with [C/H], [N/H], [Ne/H], [Ar/H], and [F/Ar] abundances. 
All the abundances of PNe in this table originate from the nebula
material. For PNe, Ar is used as metallicity indicator instead of Fe, 
because Ar is not depleted by dust. To our knowledge, BoBn 1 is 
the most metal-poor and F-rich PN among F-detected PNe. 
We also present the abundances of the CEMP star 
HE1305+0132 (HE 1305 hereafter), estimated by Schuler et al. 
(2007). It should be noted that its C and F overabundances 
are comparable to those of BoBn 1. Since the extremely enhanced 
F abundances in these objects can not be explained by 
primordial pollution of material from rapidly-rotating massive 
stars (Palacios et al. 2005), we assume that F in BoBn 1 and 
HE 1305 is mostly produced by nucleosynthesis in the 
progenitor single star or a possible binary companion.

\section{Discussion \& Conclusions}

In low-mass AGB stars (initial mass of 1--4 M$_{\odot}$; Herwig 2005) suffering the TDU and not the hot bottom
burning, the F abundance at the stellar surface is efficiently 
enhanced. Its only stable isotope, $^{19}$F, is synthesized via the 
$^{14}$N($\alpha$,$\gamma$)$^{18}$F($\beta^{+}$)$^{18}$O({\it
p},$\alpha$)$^{15}$N($\alpha$,$\gamma$)$^{19}$F reaction chain during
He-shell burning inside thermal pulses (TPs). Fluorine is further carried to
the surface of the star by the TDU, together with other 
products of He burning: mostly $^{12}$C, produced via partial He
burning, and $^{22}$Ne produced via double alpha-captures on $^{14}$N 
(Jorissen et al. 1992; Lugaro et al. 2004). The required protons are 
provided by the $^{14}$N({\it n},{\it p})$^{14}$C reaction, where the 
neutrons are produced by the $^{13}$C($\alpha$,{\it n})$^{16}$O 
reaction, which is activated at temperatures $\gtrsim$ 8.9$\times$10$^{7}$ K
(Herwig 2004). The other neutron source reaction,
$^{22}$Ne($\alpha$,{\it n})$^{25}$Mg, is not a plausible candidate to
provide neutrons for the F production because it is activated 
at higher temperatures, $\gtrsim$ 2.8$\times$10$^{9}$ K, at which 
$^{14}$N is already completely destroyed and $^{19}$F is also quickly 
consumed by $^{19}$F($\alpha$,{\it p})$^{22}$Ne reactions.

The required $^{13}$C and $^{14}$N nuclei are ingested in the convective
TPs where He burning takes place, together with the ashes
of the previous H-burning. Note that the abundances of the $^{13}$C and
$^{14}$N nuclei in the H-burning ashes have a primary component at 
low metallicity due to the fact that efficient TDU carries 
primary $^{12}$C to the stellar envelope, which is then converted into
$^{13}$C and $^{14}$N during H-burning. According to the current stellar
models, in low-mass AGB stars partial mixing of the bottom of the 
H-rich convective envelope into outermost region of the $^{12}$C-rich 
intershell layer leads to synthesis of extra $^{13}$C and $^{14}$N 
at the end of the TDU (e.g., Werner \& Herwig 2006). This is necessary to
produce enough heavy elements via the {\it s}-process in order 
to match observations of enhancements of, 
e.g., Zr, Ba, and Pb, in these stars (e.g., Busso et al. 2001).  
During the interpulse phase the $^{13}$C($\alpha$,{\it n})$^{16}$O 
reaction occurs and neutrons are released. The presence of a
partial mixing zone also contributes some extra F production. 
However, due to the uncertainties associated with the formation 
of this mixing zone (e.g., Herwig 2005), 
its contribution to the final F production should be considered 
as part of the model uncertainties (e.g., Lugaro et al. 2004; 
Karakas et al. 2007). 
Jorissen et al. (1992 for MS/S stars) and Zhang \& Liu (2005) 
showed a strong correlation between $^{12}$C and $^{19}$F 
abundances, which is expected if both $^{12}$C and $^{19}$F are carried to 
the stellar surface by the TDU. 

\begin{figure}
\epsscale{1.0}
\plotone{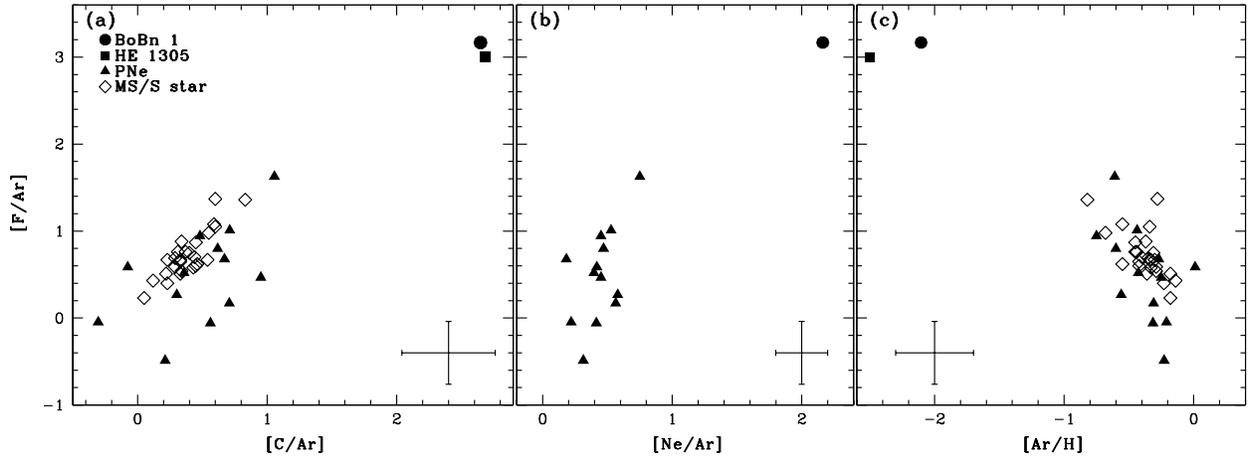}
\figcaption{
Diagrams of [F/Ar] vs. [C/Ar] ({\it left}), [Ne/Ar] ({\it middle}),
 and vs. [Ar/H] ({\it right}). For HE 1305 and MS/S stars, Fe is 
used instead of Ar as metallicity indicator. Error bars indicate 
typical inaccuracies of abundances. 
The data of MS/S stars are 
taken from Jorissen et al. (1992), Lambert et al. (1986), and 
Smith \& Lambert (1986, 1990). 
\label{diag}}
\end{figure}

In Figure \ref{diag}(a), 
we show the diagram of [C/Ar] vs. [F/Ar] for PNe, MS/S stars, and HE
1305. It clearly shows that 
[F/Ar] increases with [C/Ar]. Figure \ref{diag}(b) shows a correlation between [Ne/Ar] and [F/Ar], 
which again is to be expected if $^{22}$Ne is also carried to the 
surface by the TDU. Figure \ref{diag}(c) further shows a strong
correlation between [Ar/H] and [F/Ar]. This is predicted by the 
models because of the primary contribution to the F production 
due to the effect of the TDU, as discussed above. 

The [F/H] abundance of BoBn 1 is comparable to NGC 40, which has a
WR-type central star ([WC8]; De Marco \& Barlow 2001). The strong 
stellar wind from WR stars may also be important for the F injection. 
In BoBn 1, 
strong and narrow C~{\sc iii}, C~{\sc iv}, and N~{\sc iii} 
lines have been detected (Otsuka 2007), suggesting that the central star is 
a WELS (weak emission-line star) known as a class of H-deficient stars. 
The central star might have suffered a very late thermal 
pulse (VLTP) and become H-deficient. Zijlstra et
al. (2006) noted that the Ne enhancement of 
BoBn 1 could be caused by VLTP. The strong stellar wind 
and VLTP might contribute to the enhancements of F, C, and Ne in the nebula.

The [F/H] abundance of BoBn 1 is 
compatible with the final [F/H] predicted by the AGB single star 
models of Karakas \& Lattanzio (2007) and 
Lugaro et al. (2004) for initial masses of $\sim$2 M$_{\odot}$ stars with {\it Z} 
= 10$^{-4}$ ([Fe/H] $\simeq$ $-$2.3).
The single metal-poor star models of Fujimoto et al. (2000) and 
Suda et al. (2004) predict that stars with [Fe/H] $\lesssim$ $-$2.5 
and initial masses of 1.2--3.5 M$_{\odot}$ will become C, N, 
and {\it s}-process elements rich by the partial mixing and 
subsequent dredge-up. Such a star, however, 
would not have survived in the Galactic halo up to now. 

\begin{figure}
\epsscale{0.95}
\plotone{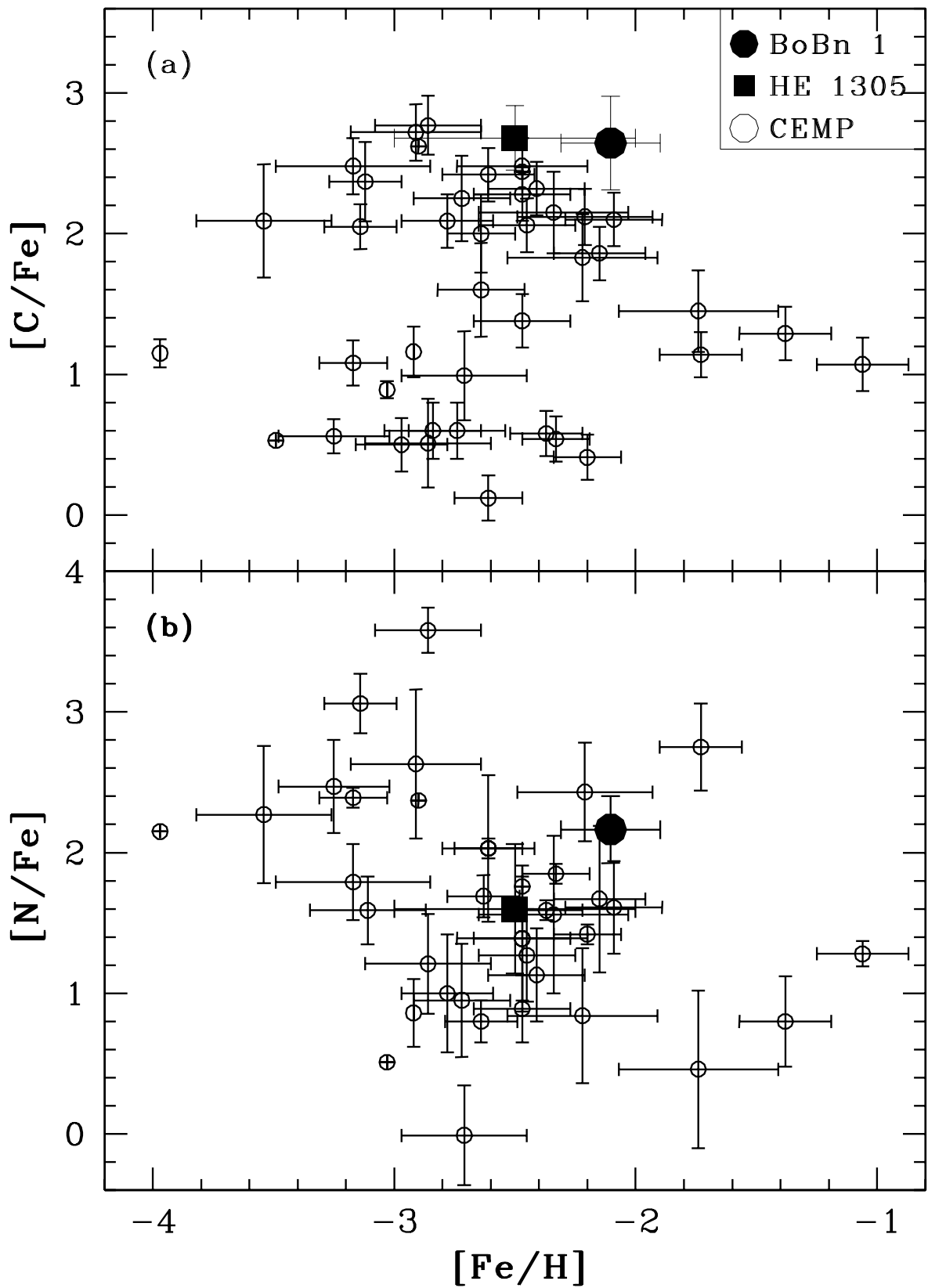}
\figcaption{
Location of BoBn 1 on the diagrams of [C/Fe] ({\it upper}) 
and [N/Fe] ({\it lower}) vs. [Fe/H], using Ar instead 
of Fe. The data of CEMP stars except HE 1305 are taken 
from Aoki et al. (2007) and Suda et al. (2008).
\label{cfe}}
\end{figure}

To circumvent this issue, 
BoBn 1 might have evolved
from a binary and have experienced binary mass transfer from a massive
companion. The abundances of BoBn 1 
are comparable to those of HE 1305 and other CEMP stars as shown in 
the diagrams of [C/Fe] and [N/Fe] versus [Fe/H] (Figure \ref{cfe}). 
As currently defined, CEMP stars fall into two categories, those that 
exhibit large enhancements of {\it s}-process elements 
(CEMP-{\it s}) and those that do not (CEMP-{\it no}) (Aoki et al. 2007).  
Most CEMP-{\it s} stars show large enhancements of C and N 
abundances. The evolution models for CEMP stars 
of Lau et al. (2007) have demonstrated that the C 
and N overabundances would be reproduced by binary interactions. 
In fact, about 70$\%$ of CEMP stars are identified as 
long-period (0.3--100 yr) binaries (Beers \& Christlieb 2005). 
Schuler et al. (2007) concluded that HE 1305 might have experienced 
mass transfer and predicted that {\it s}-process elements should 
be enhanced in this object. 
Considering the model of Karakas \& Lattanzio (2007), Lugaro et al. (2008) 
concluded that HE 1305 consists of $\sim$2 M$_{\odot}$ (primary) and 
$\sim$0.8 M$_{\odot}$ (secondary) stars with {\it Z} = 10$^{-4}$ 
and the enhanced C and F are explained by binary mass
transfer from the primary star via Roche lobe overflow and/or wind
accretion. BoBn 1 might also have evolved from a binary consisting of 
$\sim$2 M$_{\odot}$ and $\sim$0.8 M$_{\odot}$ stars, since its C and F 
abundances and metallicity are comparable to HE 1305. The enhanced C and 
F of BoBn 1 can be explained by the model of Lugaro et al. (2008) 
including the extra $^{13}$C and $^{14}$N, or the upper limit of the 
$^{18}$F($\alpha$,{\it p})$^{21}$Ne reaction. 
The chemical similarities between BoBn 1 and CEMP-{\it s} 
stars suggest that this PN shares a similar origin and evolutionary 
history, although we have not detected any {\it s}-process elements in
this object. The contradiction on the evolutionary time scale of this 
object can be avoided if BoBn 1 has indeed evolved from a CEMP-{\it s} 
star such as HE 1305. 

Abundances of {\it s}-process elements and F should be enhanced 
in K 648, H 4-1, and BoBn 1, if they have evolved from CEMP-{\it s}
stars. For K 648 and H 4-1, there have been no reports on the detection 
of F and {\it s}-process elements. High sensitivity spectroscopic 
observations are necessary in the optical and near-infrared to
investigate those elements. We would be able to settle the origin 
and evolution of halo PNe, if we find enhanced abundances for those elements.

\acknowledgements 
The authors wish to thank the anonymous referees for valuable comments. 
They wish to thank M. Fujimoto, T. Suda, M. Lugaro, and S.E. de Mink 
for helpful comments and fruitful discussion. M.O. thanks OAO staffs 
for continuous encouragements. S.H. thanks the Korea Science 
\& Engineering Foundation ARCSEC financial support.


\clearpage

\begin{table}
\caption{Fluorine ionic and elemental abundance of BoBn 1\tablenotemark{a}. \label{fl}}
\begin{tabular}{@{}lccccc@{}}
\tableline\tableline
$\lambda_{\rm lab}$ & $\lambda_{\rm obs}$ &$I$($\lambda_{\rm obs}$) &
F$^{\rm 3+}$/H$^{+}$ &icf(F) &F/H\\
(\AA)&(\AA) & ($\times$10$^{-1}$) &($\times$10$^{-8}$)&&($\times$10$^{-7}$)\\
\tableline
3996.92 &3999.76 & 1.03$\pm$0.03 & 5.08$\pm$0.21&\multicolumn{2}{c}{\nodata}\\ 
4059.90 &4062.73 & 3.09$\pm$0.04 & 5.40$\pm$0.18&\multicolumn{2}{c}{\nodata}\\
\tableline
\noalign{\smallskip}
Adopted      &\multicolumn{2}{c}{\nodata}&5.32$\pm$0.19 & 6.27$\pm$0.62 &3.33 $\pm$ 0.35\\
\tableline
\end{tabular}
\tablenotetext{a}{$I$(\hb)=100. Inaccuracies are internal only.} 
\end{table}

\begin{table}
\caption{Fluorine abundances of PNe and HE1305+0132. \label{fl2}}
\begin{tabular}{@{}lrrrrrrl@{}}
\tableline\tableline
Object &[F/H] &[C/H]&[N/H]&[Ne/H]&[Ar/H]&[F/Ar]&Ref.\\
\tableline
IC 418 & +0.41 &+0.40 &+0.09 &$-$0.09 &$-$0.27 &+0.68 &(1),(2)  \\ 
IC 2003 &+0.20 &+0.02  &$-$0.13 &$-$0.13 &$-$0.60 &+0.80 &(3)  \\ 
IC 2501 &$-$0.14 &+0.40  &+0.33 &+0.26 &$-$0.31 &+0.17 &(4),(5)  \\ 
NGC 40 &+1.02 &+0.45  &+0.10 &+0.14 &$-$0.61 &+1.63 &(6)  \\ 
NGC 2022 &+0.09 &$-$0.07  &$-$0.37 &$-$0.03 &$-$0.43 &+0.52 &(7)  \\ 
NGC 2440 &$-$0.72 &$-$0.02  &+1.17 &+0.09 &$-$0.23 &$-$0.49 &(7),(8)  \\ 
NGC 3242 &$-$0.29 &$-$0.26  &$-$0.30 &+0.02 &$-$0.56 &+0.27 &(7)  \\ 
NGC 3918 &$-$0.37 &+0.25  &+0.19 &+0.10 &$-$0.31 &$-$0.06 &(7)  \\ 
NGC 5315 &+0.60 &$-$0.06  &+0.69 &+0.43 &+0.01 &+0.59 &(7)  \\ 
NGC 6302 &$-$0.26 &$-$0.51  &+0.69 &+0.01 &$-$0.21 &$-$0.05 &(7)  \\ 
NGC 7027 &+0.22 &+0.70  &+0.31 &+0.20 &$-$0.25 &+0.47 &(9)  \\ 
NGC 4361 &+0.19 &$-$0.27  &$-$0.42&$-$0.30 &$-$0.75 &+0.94 &(10)  \\ 
NGC 7662 &+0.57 &+0.28  &+0.36&+0.09 & $-$0.44 &+1.01 &(6),(11)  \\ 
BoBn 1 &+1.06 &+0.54  &+0.06  &+0.06&$-$2.10 &+3.17 &(12)  \\ 
HE 1305&+0.50 &+0.18  &$-$0.90 &\nodata  &$-$2.50&+3.00& (13)\\
\tableline
\end{tabular}
\tablecomments{The elemental abundances in PNe except 
C of IC 2501 are estimated from forbidden lines. C of IC 2501 is 
estimated from recombination lines. For the CEMP star HE1305+0132 
(HE 1305), Fe is used as metallicity indicator. 
The solar abundances are referred to Lodders (2003).}
\tablerefs{
For column 2: see text. For columns 3--7: 
(1) Pottasch et al. (2004); 
(2) Hyung et al. (1994); 
(3) Wesson et al. (2005); 
(4) Henry et al. (2004); 
(5) Sharpee et al. (2007); 
(6) Liu et al. (2004); 
(7) Tsamis et al. (2003); 
(8) Hyung \& Aller (1998); 
(9) Zhang et al. (2005); 
(10) Torres-Peimbert et al. (1990); 
(11) Hyung \& Aller (1997);
(12) this work + Otsuka (2007); 
(13) Schuler et al. (2007).}
\end{table}

\begin{thebibliography}{}
\bibitem[]{}Acker, A. et al. 1992, Strasbourg-ESO Catalogue of Galactic
	  Planetary Nebulae 
\bibitem[Aoki et al.(2007)]{2007ApJ...655..492A} Aoki, W. et al. 
2007, \apj, 655, 492 
\bibitem{beers} Beers, T.~C. \& Christlieb, N.\ 2005, ARAA, 43, 531
\bibitem[Busso et al.(2001)]{2001ApJ...557..802B} Busso, M. et al. 
2001, \apj, 557, 802 
\bibitem[Dekker et al.(2000)]{2000SPIE.4008..534D} 
Dekker, H. et al. 2000, \procspie, 4008, 534
\bibitem[De Marco \& Barlow(2001)]{2001Ap&SS.275...53D} De Marco, O. \&
	  Barlow, M.~J. 2001, \apss, 275, 53 
\bibitem[Fujimoto et al.(2000)]{2000ApJ...529L..25F} Fujimoto, M.~Y., 
Ikeda, Y., \& Iben, I.~J. 2000, \apjl, 529, L25 
\bibitem[Garstang(1951)]{1951MNRAS.111..115G} Garstang, R.~H. 1951, 
\mnras, 111, 115 
\bibitem[Henry et al.(2004)]{2004AJ....127.2284H} Henry, R.~B.~C., Kwitter, 
K.~B., \& Balick, B. 2004, \aj, 127, 2284 
\bibitem[Herwig(2004)]{2004ApJ...605..425H} Herwig, F. 2004, \apj, 605, 
425 
\bibitem[Herwig(2005)]{2005ARA&A..43..435H} Herwig, F. 2005, \araa, 43, 435 
\bibitem[Howard et al.(1997)]{1997MNRAS.284..465H} Howard, J.~W., 
Henry, R.~B.~C., \& McCartney, S. 1997, \mnras, 284, 465 
\bibitem[Hyung et al.(1994)]{1994PASP..106..745H} Hyung, S., Aller, L.~H., 
\& Feibelman, W.~A. 1994, \pasp, 106, 745 
\bibitem[Hyung \& Aller(1997)]{1997ApJ...491..242H} Hyung, S. \& Aller, 
L.~H. 1997, \apj, 491, 242 
\bibitem[Hyung \& Aller(1998)]{1998PASP..110..466H} Hyung, S. \& Aller, 
L.~H. 1998, \pasp, 110, 466  
\bibitem[Jorissen et al.(1992)]{1992A&A...261..164J} Jorissen, A., Smith, 
V.~V., \& Lambert, D.~L. 1992, \aap, 261, 164 
\bibitem[Karakas \& Lattanzio(2007)]{2007PASA...24..103K} Karakas, A. \& 
Lattanzio, J.~C. 2007, PASA, 24, 103 
\bibitem[Karakas et al.(2007)]{2007arXiv0712.2883K} Karakas, A. et al. 
2007, ArXiv e-prints, 712, arXiv:0712.2883 
\bibitem[]{} Kingsburgh, R.~L. \& Barlow, M.~J. 1992, \mnras, 257, 317
\bibitem[Lambert et al.(1986)]{1986ApJS...62..373L} Lambert, D.~L., Gustafsson, B., Eriksson, K., \& Hinkle, K.~H. 1986, \apjs, 62, 373 
\bibitem{lau} Lau, H.~H.~B., Stancliffe, R.~J., \& Tout, C.~A. 2007, 
	\mnras, 378, 563
\bibitem[Lennon \& Burke(1994)]{1994A&AS..103..273L} Lennon, D.~J. \& 
Burke, V.~M. 1994, \aaps, 103, 273 
\bibitem[Liu et al.(2004)]{2004MNRAS.353.1231L}
Liu, Y., Liu, X.-W., Barlow, M.~J., \& Luo, S.-G. 2004, \mnras, 353, 1251 
\bibitem[Lodders(2003)]{2003ApJ...591.1220L} Lodders, K. 2003, \apj, 591, 
1220 
\bibitem[Lugaro et al.(2004)]{2004ApJ...615..934L} 
Lugaro, M. et al. 2004, \apj, 615, 934 
\bibitem[Lugaro et al.(2008)]{} Lugaro, M. et al. 2008, \aap, 484, L27
\bibitem[]{}Otsuka, M. 2007, Ph.D. thesis, Tohoku Univ. (Japan) 
\bibitem[]{}Pereira, C.-B. \& Miranda, L.-F. 2007, \aap, 467, 1249
\bibitem[Pottasch et al.(2004)]{2004A&A...423..593P} Pottasch, S.~R., 
Bernard-Salas, J., Beintema, D.~A., \& Feibelman, W.~A. 2004, \aap, 423, 593 
\bibitem[Schuler et al.(2007)]{} Schuler, S.~C. et al. 2007, \apjl, 667, L81 
\bibitem[Sharpee et al.(2007)]{2007ApJ...659.1265S} Sharpee, B. et al. 2007, \apj, 659, 1265 
\bibitem[Smith \& Lambert(1986)]{1986ApJ...311..843S} Smith, V.~V. \& 
Lambert, D.~L. 1986, \apj, 311, 843 
\bibitem[Smith \& Lambert(1990)]{1990ApJS...72..387S} Smith, V.~V. \& 
Lambert, D.~L. 1990, \apjs, 72, 387
\bibitem[Storey \& Zeippen(2000)]{2000MNRAS.312..813S} Storey, P.~J. \&
	  Zeippen, C.~J. 2000, \mnras, 312, 813
\bibitem[Straniero et al.(2006)]{2006NuPhA.777..311S} Straniero, O., 
Gallino, R., \& Cristallo, S. 2006, Nuclear Physics A, 777, 311 
\bibitem[Suda et al.(2004)]{2004ApJ...611..476S} Suda, T. et al. 
2004, \apj, 611, 476 
\bibitem[Suda et al.(2008)]{} Suda, T. et al. 2008, arXiv:0806.3697
\bibitem[Torres-Peimbert et al.(1990)]{1990A&A...233..540T} 
Torres-Peimbert, S., Peimbert, M., \& Pena, M. 1990, \aap, 233, 540 
\bibitem[Tsamis et al.(2003)]{2003MNRAS.345..186T} Tsamis, Y.~G. et al. 
2003, \mnras, 345, 186 
\bibitem[Werner \& Herwig(2006)]{2006PASP..118..183W} Werner, K. \& 
Herwig, F. 2006, \pasp, 118, 183 
\bibitem[Wesson et al.(2005)]{2005MNRAS.362..424W} Wesson, R., Liu, X.-W., 
\& Barlow, M.~J. 2005, \mnras, 362, 424 
\bibitem[Zhang \& Liu(2005)]{2005ApJ...631L..61Z} Zhang, Y. \& Liu,
	  X.-W. 2005, \apjl, 631, L61 
\bibitem[Zhang et al.(2005)]{2005A&A...442..249Z} Zhang, Y. et al. 2005, \aap, 442, 249
\bibitem[]{} Zijlstra, A.~A. et al. 2006, \mnras, 369, 875
\end{thebibliography}
\end{document}